\documentclass[sigconf]{acmart}

\usepackage{multirow}
\usepackage{makecell}
\usepackage{enumitem}
\usepackage{siunitx}
\usepackage{xspace}
\usepackage{subcaption}
\usepackage[ruled,linesnumbered,vlined]{algorithm2e}
\usepackage{soul}
\usepackage[]{collab}

\let\oldnl\nl
\newcommand{\nonl}{\renewcommand{\nl}{\let\nl\oldnl}}
\SetAlFnt{\small}

\DeclareMathOperator*{\argmin}{arg\,min}  

\collabAuthor{zb}{teal}{Zhuangbin Chen}

\AtBeginDocument{%
  }

\copyrightyear{2026}
\acmYear{2026}
\setcopyright{cc}
\setcctype{by-nc-nd}
\acmConference[ICSE '26]{2026 IEEE/ACM 48th International Conference on Software Engineering}{April 12--18, 2026}{Rio de Janeiro, Brazil}
\acmBooktitle{2026 IEEE/ACM 48th International Conference on Software Engineering (ICSE '26), April 12--18, 2026, Rio de Janeiro, Brazil}
\acmPrice{}
\acmDOI{10.1145/3744916.3773239}
\acmISBN{979-8-4007-2025-3/2026/04}

\newcommand\alias{\textsc{Metronome}\xspace}

\newsavebox{\obsbox} 
\newenvironment{observationbox}{%
  \par\vspace{6pt}\noindent 
  \setlength{\fboxrule}{0.2pt}%
  \setlength{\fboxsep}{3pt}%
  \begin{lrbox}{\obsbox}%
    \begin{minipage}{\dimexpr\linewidth-2\fboxsep-2\fboxrule\relax}
    \vspace{1pt}%
}{%
    \end{minipage}%
  \end{lrbox}%
  \fcolorbox{black}{white!90!black}{\usebox{\obsbox}}%
  \par\vspace{6pt} 
}

\begin{document}

\title{\alias: Differentiated Delay Scheduling for Serverless Functions}


\author{Zhuangbin Chen}
\affiliation{%
  \institution{Sun Yat-sen University}
  \city{Zhuhai}
  \country{China}}
\email{chenzhb36@mail.sysu.edu.cn}
\orcid{0000-0001-5158-6716}

\author{Juzheng Zheng}
\affiliation{%
  \institution{Sun Yat-sen University}
  \city{Zhuhai}
  \country{China}}
\email{zhengjzh8@mail2.sysu.edu.cn}
\orcid{0009-0007-2084-3485}

\author{Zibin Zheng}
\authornote{Zibin Zheng is the corresponding author.}
\affiliation{%
  \institution{Sun Yat-sen University}
  \city{Zhuhai}
  \country{China}}
\email{zhzibin@mail.sysu.edu.cn}
\orcid{0000-0001-7872-7718}

\renewcommand{\shortauthors}{Chen, et al.}

\begin{abstract}


Function-as-a-Service (FaaS) computing is an emerging cloud computing paradigm for its ease-of-management and elasticity.
However, optimizing scheduling for serverless functions remains challenging due to their dynamic and event-driven nature.
While data locality has been proven effective in traditional cluster computing systems through delay scheduling, its application in serverless platforms remains largely unexplored.
In this paper, we systematically evaluate existing delay scheduling methods in serverless environments and identify three key observations: 1) delay scheduling benefits vary significantly based on function input characteristics; 2) serverless computing exhibits more complex locality patterns than cluster computing systems, encompassing both data locality and infrastructure locality; and 3) heterogeneous function execution times make rule-based delay thresholds ineffective. Based on these insights, we propose \alias, a differentiated delay scheduling framework that employs predictive mechanisms to identify optimal locality-aware nodes for individual functions. \alias leverages an online Random Forest Regression model to forecast function execution times across various nodes, enabling informed delay decisions while preventing SLA violations. Our implementation on OpenLambda shows that \alias significantly outperforms baselines, achieving 64.88\%--95.83\% reduction in mean execution time for functions, while maintaining performance advantages under increased concurrency levels and ensuring SLA compliance.
\end{abstract}





\maketitle

\section{Introduction}


Function-as-a-Service (FaaS) computing offers key benefits including reduced operational overhead, automatic scalability, and cost efficiency via pay-per-use pricing~\cite{crane2017exploration,mcgrath2017serverless,baldini2017serverless}.
However, the dynamic and event-driven nature of serverless workloads presents unique challenges for their scheduling~\cite{shafiei2022serverless,li2022serverless}.
Recent studies~\cite{DBLP:conf/eurosys/AbdiGLFCGBBF23,carver2020wukong,DBLP:conf/usenix/MahgoubSMKCB21,Klimovic_Wang_Stuedi_Trivedi_Pfefferle_Kozyrakis_2018} have shown that \textit{data locality} between data-intensive functions plays a critical role in optimizing resource utilization.
For example, routing subsequent requests from the same user to the same function instance can significantly improve cache effectiveness, while scheduling data analytics tasks that consume each other's outputs on the same instance can minimize network data transfer overhead.

In traditional cluster computing systems such as Hadoop and Dryad, previous work exploits and enforces data locality through the strategy of \textit{delay scheduling}~\cite{DBLP:conf/eurosys/ZahariaBSESS10}.
The principle of delay scheduling involves intentionally postponing the execution of tasks until nodes that contain the required data become available.
Specifically, when a node requests a task, if the head-of-line job cannot launch a local task, the scheduler will skip it and examine the subsequent jobs.
If a job has been skipped long enough, it will be permitted to launch non-local tasks to avoid starvation.
The key insight is that tasks typically complete quickly, so the local nodes with the required data are likely to be freed up within a few seconds.
This technique has been shown to significantly minimize data transfer overhead and improve response times.
Similar to the tasks in cluster computing systems, serverless functions also benefit from data locality and typically have short execution times (ranging from milliseconds to a few seconds).
While Meta's XFaaS platform~\cite{DBLP:conf/osdi/SchusterDVDHGKMAA21} employs ``time-shifting'' techniques to defer delay-tolerant functions to off-peak hours, the primary objective is capacity management and cost reduction, which require developers to manually annotate functions as non-critical or specify execution deadlines.
In contrast, effective locality-aware scheduling requires automated, real-time decisions.






To fill this significant gap, in this work we systematically evaluate the performance of existing delay scheduling methods~\cite{DBLP:conf/osdi/SchusterDVDHGKMAA21,DBLP:conf/eurosys/ZahariaBSESS10,Ousterhout_Wendell_Zaharia_Stoica_2013,wang2019pigeon} in serverless environments.
Specifically, we deploy four widely-used serverless applications from FunctionBench~\cite{FunctionBench} and generate workloads based on production traffic~\cite{HuaweiWorkload}.
Through extensive experiments, we find that the unique computing paradigm and intricate execution patterns of serverless functions render existing approaches ineffective.
Particularly, we have identified three key observations that serve as essential design principles for optimizing delay scheduling in serverless environments:

\begin{itemize}[noitemsep,leftmargin=5.5mm]

    \item \textit{Delay scheduling is not universally beneficial, depending on the input characteristics of individual functions.}
    Specifically, functions with large input data sizes benefit significantly from data locality, as it minimizes substantial data transfer overhead, while those with smaller inputs show minimal performance gains.
    Existing work fails to account for this critical factor, leaving a gap in identifying which functions are truly suitable candidates for delay scheduling techniques.
    

    \item \textit{The locality characteristics in serverless computing exhibit greater complexity compared to traditional systems.}
    Unlike traditional cluster computing systems where data locality is the primary concern, we identify another critical type of locality, i.e., \textit{infrastructure locality}, which refers to the reuse of warm containers and cached dependencies across function invocations on the same node.
    Moreover, functions with different characteristics benefit from different forms of locality.


    \item \textit{The heterogeneity in function execution times makes it challenging to determine the appropriate delay thresholds.}
    While most serverless functions have short execution times, there exists a non-negligible portion of long-running functions that can take tens of seconds to minutes to complete.
    Employing a rule-based delay strategy without considering these variations could result in SLA violations, particularly for time-sensitive functions.
\end{itemize}

To address these limitations, we propose \alias, which employs differentiated delay scheduling based on specific function characteristics, in contrast to existing one-size-fits-all solutions.
At its core, \alias leverages predictive delay mechanisms to identify optimal locality-aware nodes (data versus infrastructure) for individual functions (with varying input data sizes and dependency requirements).
Specifically, \alias employs an online Random Forest Regression (RFR) model that forecasts function execution times across various nodes based on comprehensive feature sets, including function attributes, node metrics, and historical execution data.
This not only enables us to compare the performance benefits of different delay decisions but also helps determine the maximum allowable delay duration to prevent SLA violations.
The online learning capability of the prediction model also facilitates it to rapidly adapt to new function characteristics and system dynamics.
We have implemented \alias on OpenLambda~\cite{DBLP:conf/usenix/OakesYZHHAA18,hendrickson2016serverless} and compared its performance against traditional delay scheduling approaches.
The results demonstrate that \alias can significantly improve the performance of function execution over baselines, while avoiding SLA violations.


The major contributions of this work are summarized as follows:


\begin{itemize}[noitemsep,topsep=0pt,leftmargin=5.5mm]
    \item A comprehensive study that reveals the limitations of existing static, rule-based delay scheduling techniques in serverless environments, which establishes essential design principles for optimizing delay scheduling for serverless functions.

    \item A differentiated delay scheduling framework that determines whether a function invocation should be delayed and the optimal locality type.
    This is achieved by an execution time prediction model with online learning capabilities.
    It is able to adaptively adjust delay thresholds to 
    optimize locality benefits while ensuring SLA compliance.


    \item A practical implementation\footnote{\url{https://github.com/OpsPAI/Metronome}} on OpenLambda~\cite{DBLP:conf/usenix/OakesYZHHAA18,hendrickson2016serverless} that leverages gRPC and server-side streaming for efficient communication between components, enabling asynchronous model inference, online learning, and system metrics collection through server push notifications.
    These operations are offloaded from the critical path of function execution while maintaining low-latency scheduling and prediction accuracy.
\end{itemize}


\section{Background and Motivation}
\label{sec:background_and_motivation}

\subsection{Delay Scheduling}

Delay scheduling was first introduced by Zaharia et al.~\cite{DBLP:conf/eurosys/ZahariaBSESS10} to address the conflict between data locality and scheduling fairness in traditional big data processing systems like Hadoop.
The core idea is to temporarily delay a task's scheduling when its preferred node (typically the one containing its input data) is unavailable, allowing other tasks to be executed first.
Such an approach has proven highly effective in cluster computing environments, where data transfer costs often outweigh waiting time.
In particular, this simple mechanism is able to achieve nearly optimal data locality while maintaining fair sharing in Hadoop clusters.


This delay scheduling strategy exhibits promising potential in serverless computing environments for several reasons.
First, the majority of serverless functions have short execution times (typically milliseconds to seconds), which means even brief scheduling delays could create sufficient opportunities for better locality without significantly impacting overall latency.
Second, serverless platforms often manifest high concurrency with multiple function instances competing for resources, making it feasible to find alternative tasks to execute during delay periods.
Third, the pay-per-use pricing model of serverless computing makes performance optimization particularly valuable, as even small improvements in execution time can lead to significant cost savings at scale.

However, adapting delay scheduling to serverless environments presents unique challenges that warrant investigation.
The fundamental differences between traditional batch processing systems and serverless platforms, such as the presence of cold starts, strict resource isolation requirements, and highly variable function characteristics, raise questions about the direct applicability of the original delay scheduling approach.
To better understand these challenges and evaluate the potential benefits, we conduct a series of experiments examining how delay scheduling performs in serverless scenarios with different workload characteristics.

\subsection{An Empirical Study of Delay Scheduling in Serverless Functions}
\label{sec:empirical_study}

To understand the effectiveness and limitations of traditional delay scheduling in serverless environments, we conduct a comprehensive study using four representative serverless applications from FunctionBench~\cite{FunctionBench}, a widely-used benchmark suite~\cite{fuerst2021faascache,zhang2021faster,ustiugov2021benchmarking}.
Table~\ref{tab:function-characteristics} provides a detailed overview, including the functions of each application, the number of their dependencies, and descriptions.

\begin{table}[t]
    \caption{Descriptions of benchmark functions}
    \Description{Tabular summary listing four serverless applications, their constituent functions, the number of library or service dependencies for each function, and a brief textual description of what each function does.}
    \label{tab:function-characteristics}
    \vspace{-2pt}
    \resizebox{\columnwidth}{!}{%
    \begin{tabular}{llcp{3.3cm}}
    \specialrule{0.35mm}{0em}{0em}
    \textbf{Application} & \textbf{Functions} & \textbf{Deps Count} & \textbf{Description} \\
    \specialrule{0.15mm}{0em}{0em}
    \specialrule{0.15mm}{.1em}{0em}
    \multirow{3}{*}{\makecell{Video\\Processing}} 
     & video-split & 5 & Splits video into segments \\
     & video-transcode & 4 & Transcodes segments \\
     & video-merge & 4 & Merges segments \\
    \midrule
    \multirow{3}{*}{\makecell{Log\\Analysis}} 
     & log-split & 0 & Splits log files \\
     & log-analyze & 3 & Computes IP frequencies \\
     & log-merge & 2 & Aggregates results \\
    \midrule
    \multirow{2}{*}{\makecell{Doc\\Conversion}} 
     & any2md-validate & 2 & Format validation \& prep. \\
     & any2md-process & 34 & Doc to markdown conv. \\
    \midrule
    \multirow{2}{*}{\makecell{ML\\Inference}} 
     & ml-normalize & 4 & Data normalization \\
     & ml-process & 17 & Model prediction \\
    \specialrule{0.35mm}{0em}{0em}
    \end{tabular}%
    }
    \vspace{-2pt}
\end{table}

\begin{itemize}[noitemsep,leftmargin=5.5mm]
    \item \textit{Video Processing}: An application that splits input videos, performs parallel transcoding of video segments, and merges the processed segments. Each step generates intermediate results that need to be transferred between functions.
    
    \item \textit{Log Analysis}: A service that splits log files, performs parallel analysis to compute per-IP access frequencies from log segments, and aggregates the results.
    The workflow typically processes log files and generates intermediate analysis results between processing stages.
    
    \item \textit{Document Conversion}: A function that transforms various document formats to markdown. While the input and output files are relatively small, the function requires several document processing libraries and runtime dependencies.
    
    \item \textit{ML inference}: A service that performs model-based prediction.
    It depends on multiple ML frameworks that need to be loaded into the runtime environment.
\end{itemize}

These applications are carefully selected to represent common serverless workload patterns observed in production scenarios, i.e., data processing pipelines (video processing and log analysis), standalone compute-intensive tasks (document conversion), and machine learning services (ML inference).
Moreover, they exhibit diverse characteristics in terms of execution duration (ranging from milliseconds to minutes), resource requirements (CPU, memory, and I/O patterns), and dependency complexity (from lightweight libraries to heavy ML frameworks), making them ideal for evaluating scheduling strategies under different scenarios.

To ensure a realistic evaluation, we generate workload patterns based on production cloud serverless traces~\cite{HuaweiWorkload}, which contain over 1.4 trillion invocations across more than 5,000 functions. Our workload generation process systematically extracts and synthesizes key characteristics from production environments. Specifically, we analyze the traces to extract key metrics including request volumes, invocation frequencies, inter-arrival time distributions, and function execution patterns. We calculate statistical properties such as arrival rate distributions (from 0.1 to 50 requests per second), function popularity patterns (following power-law distributions), and temporal correlations between dependent functions in workflows. The workload generator then synthesizes these characteristics to create realistic request streams that preserve the essential properties of production workloads while adapting to our experimental applications.
For each application, we conduct 100 independent end-to-end executions and record their execution times.

In our experiments, we compare two scheduling strategies:

\begin{itemize}[noitemsep,leftmargin=5.5mm]
    \item \textit{Basic Data-locality Scheduling}: A basic locality-aware scheduling strategy that considers data locality but without delay mechanisms. When a function reaches the head of the scheduling queue, the scheduler first checks available nodes that contain relevant input data (i.e., nodes where the predecessor functions in the workflow were executed).
    If such nodes exist, the scheduler selects one based on current load conditions.
    If no nodes with data locality are available, the function is scheduled immediately to non-local nodes.
    
    \item \textit{Rule-based Delay Scheduling}: An implementation of traditional delay scheduling~\cite{DBLP:conf/eurosys/ZahariaBSESS10,Ousterhout_Wendell_Zaharia_Stoica_2013,wang2019pigeon}.
    When a function reaches the head of the queue, if no data-locality nodes are available, instead of immediately scheduling to any available node, the function is marked for delay and remains in the queue. The scheduler records the delay start time and continues to check for nodes with data locality whenever resources are released.
    The delay timeout is calculated based on our serverless environment characteristics and workload patterns, following the rules in~\cite{DBLP:conf/eurosys/ZahariaBSESS10}. If the delay exceeds this calculated timeout, the function is scheduled to the next available node regardless of locality.
\end{itemize}

\begin{figure}[t]
    \centering
    \includegraphics[width=0.9\columnwidth]{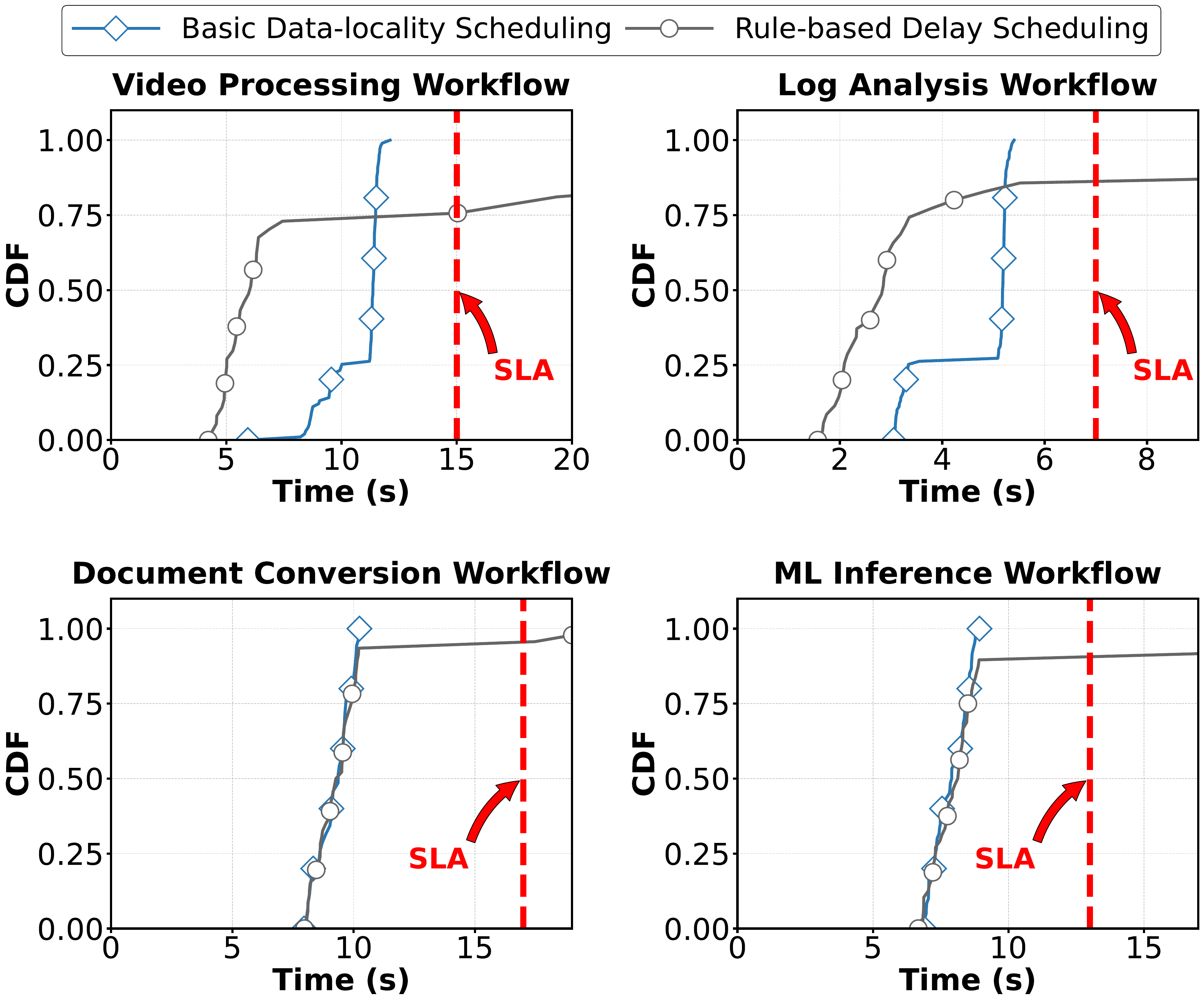}
    \vspace{-2pt}
    \caption{CDF of function execution time}
    \vspace{-3pt}
    \label{fig:execution_cdfs}
\end{figure}

The results reveal interesting and sometimes counter-intuitive patterns, as illustrated in Figure~\ref{fig:execution_cdfs}.
The cumulative distribution function (CDF) shows the proportion of function runs with an execution time less than or equal to a given value.
A curve that rises more steeply and is shifted to the left indicates better performance.
The video processing shows mixed performance results with a modest reduction (3.9\%) in mean execution time but a more substantial reduction (47.1\%) in median time.
Similarly, the log analysis achieves a 2.7\% reduction in mean execution time and a 45.3\% in median time.
However, this comes at the cost of significantly degraded tail latencies, i.e., the video processing's 95th and 99th percentile latencies increase by 142.9\% and 151.7\%, respectively, while the log analysis shows even more severe degradation with increases of 202.6\% and 228.1\%. 
Approximately 10.2\% and 9.8\% of function invocations result in SLA violations for video and log workflows, respectively.
For the other two functions, the results are consistently negative.
The document conversion shows increased latencies across all metrics, i.e., 8.5\% higher mean, 2.1\% higher median, and 51.3\% higher 95th percentile, with 6\% SLA violations. 
The ML inference exhibits even more severe degradation, i.e., 17.6\% higher mean, 4.4\% higher median, and 118.4\% higher 95th percentile, with 10\% SLA violations.


While we observe some potential benefits in median cases, the significant tail latency degradation and SLA violations across all function types indicate that a simple delay scheduling is insufficient.
To better understand the results, we conduct a deeper analysis of the underlying patterns and identify three key observations.


\begin{observationbox}
\textbf{Observation 1:}
    The effectiveness of data-locality-based delay scheduling is highly dependent on the input data size.
    Data-intensive workloads tend to benefit from data locality optimization, while smaller workloads often demonstrate minimal improvement or even performance degradation.
\end{observationbox}



Our results show that the effectiveness of data-locality scheduling directly correlates with function input size.
Workflows such as video processing and log analysis generally involve larger input data sizes, with median sizes ranging from tens of megabytes to several gigabytes. 
These workflows exhibit notable benefits from data locality optimization.
In contrast, functions like document conversion and ML inference typically process smaller inputs, with median sizes of only a few megabytes.
For these functions, delay scheduling often yields near-zero improvement or even performance degradation.
This pattern persists even in outlier scenarios.
When video and log processing functions occasionally handle smaller input sizes, they also gain minimal benefits from delay scheduling, similar to the other applications.
This consistent correlation between input size and scheduling effectiveness confirms that when data movement overhead is small, the potential benefits from data locality optimization become negligible or are outweighed by other factors.
That is, the delay scheduling introduces unnecessary waiting time without compensating benefits in reduced data transfer overhead.

\begin{figure}[t]
    \centering
    \includegraphics[width=0.9\columnwidth]{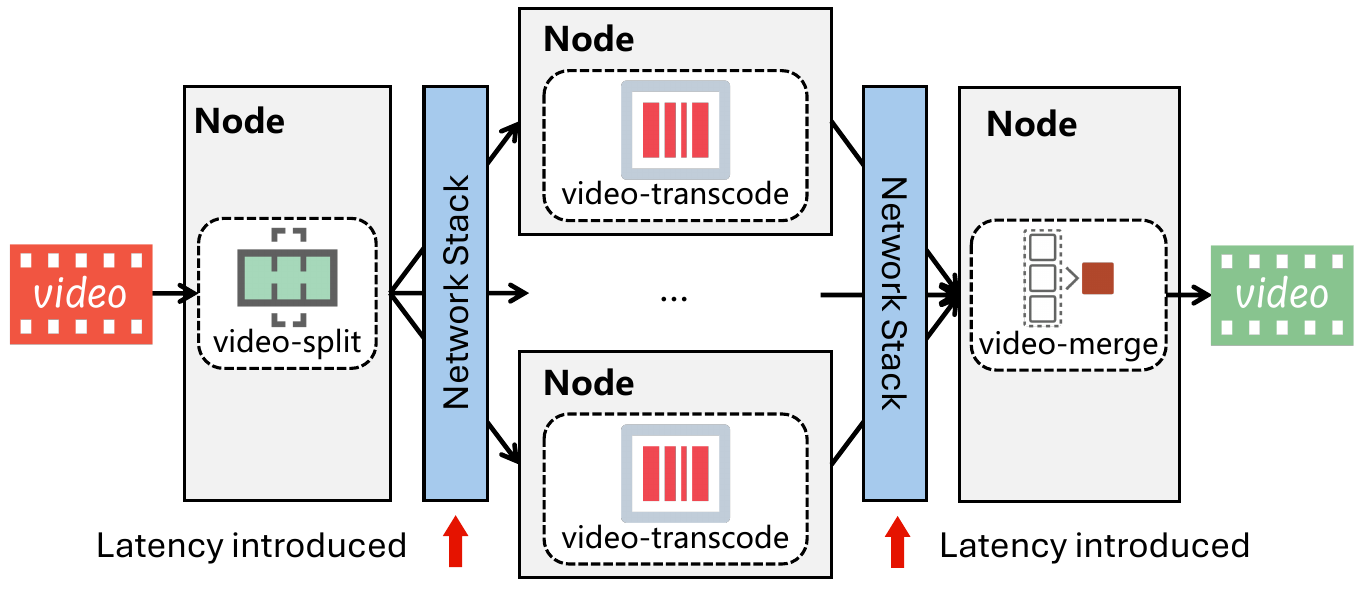}
    \vspace{-5pt}
    \caption{Video processing workflow without data locality}
    \Description{Diagram of a video processing workflow whose functions are placed on different nodes connected by arrows, so that intermediate data must traverse the network between machines.}
    \label{fig:video-workflow-without}
    \vspace{-3pt}
\end{figure}

\begin{figure}[t]
    \centering
    \includegraphics[width=0.84\columnwidth]{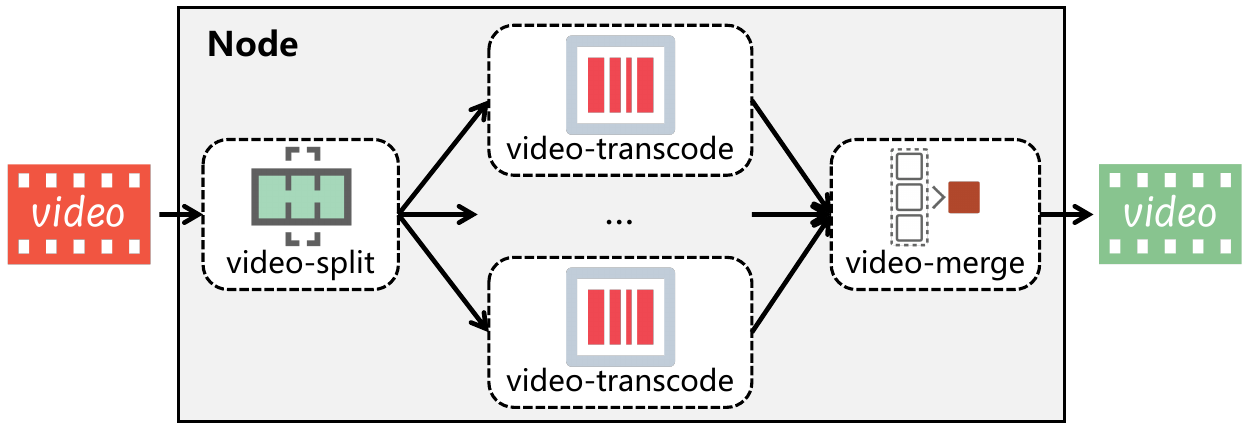}
    \vspace{-5pt}
    \caption{Video processing workflow with data locality}
    \Description{Diagram of the same video processing workflow with all functions co-located on a single node, with arrows indicating local communication paths that avoid cross-machine data transfers.}
    \label{fig:video-workflow-with}
    \vspace{-3pt}
\end{figure}

The observed correlation between input data size and delay scheduling effectiveness can be explained by examining the underlying data movement mechanisms.
As illustrated in Figure~\ref{fig:video-workflow-without}, when functions are scheduled across different nodes, data transfer between workflow stages must occur over the network, introducing significant latency overhead for intermediate results.
This can become a major performance bottleneck, especially for data-intensive applications.
In contrast, Figure~\ref{fig:video-workflow-with} demonstrates how co-locating functions on the same node minimizes data transfer overhead through efficient local communication channels such as shared memory~\cite{DBLP:conf/usenix/MahgoubSMKCB21,DBLP:conf/eurosys/LuWH0W024}, Unix domain sockets~\cite{DBLP:conf/sigcomm/LiCWBZ19,DBLP:conf/asplos/JiaW21}, eBPF-based architectures~\cite{DBLP:conf/sigcomm/QiMZWR22}, and local filesystem access~\cite{DBLP:conf/asplos/JiaW21}. This optimization is particularly beneficial for applications that process and transfer substantial amounts of intermediate data.

Based on these observations, we conclude that serverless functions may not benefit from data locality-aware scheduling in all cases.
Therefore, before applying delay scheduling, we need to carefully assess whether it would actually benefit the function's performance, given its characteristics and current system conditions.
This evaluation requires considering factors like input data size and potential performance gains from data locality.



\begin{observationbox}
\textbf{Observation 2:}
    In serverless computing, while data locality remains crucial for workflow-based applications, infrastructure locality emerges as another critical dimension due to the container-based execution model and package dependency management.
\end{observationbox}

Our experiments with document conversion and ML inference reveal a pattern that challenges the traditional focus on data locality in distributed systems.
Despite having minimal data transfer requirements, these functions consistently show degraded performance under data locality-focused delay strategies.
Further investigation reveals a critical dimension of locality unique to serverless computing.
The performance degradation can be attributed to the container-based execution model and complex package dependency management inherent in serverless platforms.
The document conversion, requiring 34 packages, shows increased latencies across all metrics, while the ML inference, with 17 dependencies, exhibits even more severe degradation. When these dependency-heavy functions are scheduled purely based on data locality considerations, they may miss opportunities to reuse existing infrastructure components that could significantly improve their performance.

This infrastructure reuse (which we term infrastructure locality) can manifest in multiple forms in serverless environments, e.g., warm containers, zygote processes~\cite{Catalyzer,DBLP:conf/usenix/LiG0CXZSMY0G22}, and pre-installed package dependencies.
Previous studies like SOCK~\cite{DBLP:conf/usenix/OakesYZHHAA18} and RainbowCake~\cite{DBLP:conf/asplos/YuRFTLZWP24} have demonstrated that serverless functions can benefit from different levels of such infrastructure reuse.
The most beneficial form is the reuse of warm containers that have already been initialized with the function code and runtime environments~\cite{Medes,Pronghorn,shillaker2020faasm,shen2021defuse}.
When warm containers are not available, nodes with cached package dependencies can still offer significant benefits by avoiding time-consuming package installations.

The importance of infrastructure reuse is particularly evident in functions like document conversion and ML inference, where the overhead of container initialization and package importing can dominate the execution time.
These functions, with their numerous package dependencies but relatively small data transfer requirements, demonstrate that optimal scheduling decisions in serverless environments must consider more than just data placement.
In contrast, video processing and log analysis functions, which have significantly fewer dependencies in our experimental setup (as shown in Table~\ref{tab:function-characteristics}), do not suffer the same performance degradation under delay scheduling, even when processing smaller input sizes.

\begin{observationbox}
\textbf{Observation 3:} 
    Function execution times exhibit high variability and heterogeneity, rendering rule-based delay thresholds impractical.
    Optimal delay decisions must consider both current system conditions and historical execution patterns, and should be dynamically determined in a predictive way.
\end{observationbox}

Our results with rule-based delay scheduling reveal significant challenges in determining appropriate delay thresholds.
The execution times of our benchmark functions are highly variable due to multiple dynamic factors. Node-level conditions such as CPU utilization, memory pressure, and I/O contention significantly impact function performance.
Function-specific characteristics like input data size, container warmth state, and package dependency availability introduce further variability, as evident from the previous two observations. 
For instance, the video processing application's execution time could vary by up to 151.7\% at the 99th percentile, while the ML inference service shows even greater variation of up to 174.9\%.
Furthermore, our experiments show that the time required for target nodes to free up resources is also highly unpredictable.
While nodes often release resources quickly, improving median execution times, excessive waiting times can occur. This results in substantial SLA violations, i.e., 6\% for document conversion and around 10\% for other applications.
This is because the heuristic delay strategy waits for an inappropriate time for preferred nodes to become available, without considering either the variable function execution times or the unpredictable resource release patterns.


This dual variability makes it impractical to use traditional, rule-based delay thresholds based on the simple setting in~\cite{DBLP:conf/eurosys/ZahariaBSESS10}.
This calls for an adaptive delay duration that needs to be dynamically determined based on both factors, i.e., how long the function is likely to take to execute (which varies based on current system conditions and function characteristics) and how long we might need to wait for resources to become available on the preferred nodes. 
A heuristic threshold~\cite{DBLP:conf/eurosys/ZahariaBSESS10} cannot capture this complex interplay of factors and may lead to either pre-mature scheduling (missing locality benefits) or excessive delays (causing SLA violations).

\begin{figure}
\centering
\includegraphics[width=0.94\columnwidth]{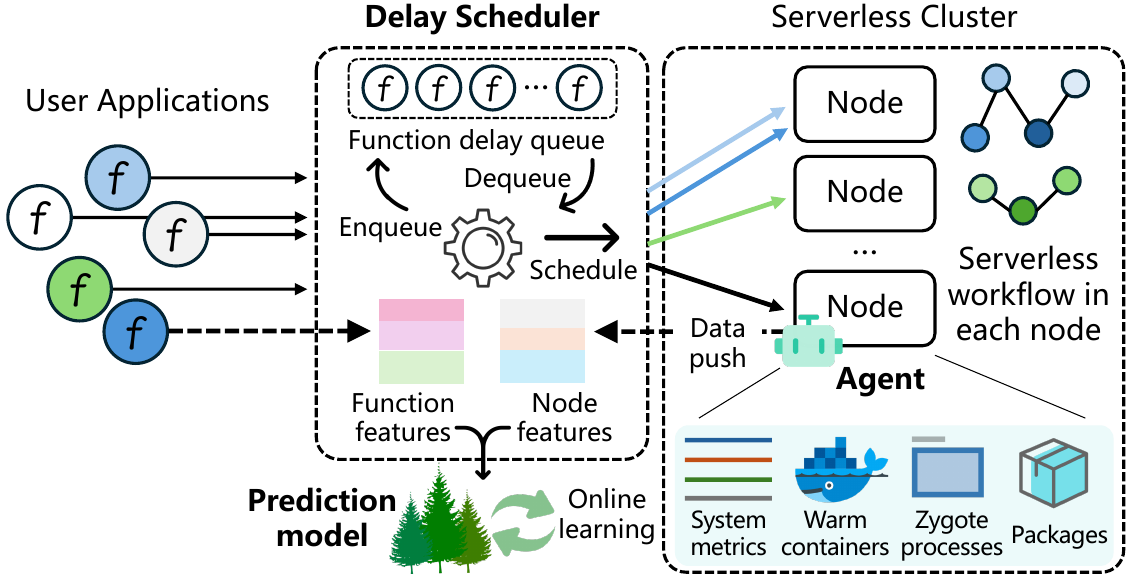}
\vspace{-3pt}
\caption{Overall architecture of \alias}
\Description{The diagram of \alias architecture with boxes for components such as clients, scheduler, learning models, and worker nodes, connected by arrows that indicate the directions of control and data flow.}
\label{fig:design-arch}
\vspace{-5pt}
\end{figure}


\section{Methodology}

\subsection{Overview}

Based on our observations, we identify three key design principles.
First, effective delay scheduling should consider the unique input characteristics, as smaller inputs may compromise data-locality benefits.
Second, functions exhibit varying sensitivities to different locality types, i.e., functions that transfer large data volumes benefit from data locality, whereas dependency-heavy functions gain more from infrastructure locality.
Third, the substantial heterogeneity in function execution times renders rule-based delay thresholds ineffective.
We need to balance the tradeoff between waiting for preferred nodes and meeting SLA compliance.

Building on these insights, we propose \alias, a differentiated scheduling framework that uses per-function execution time predictions to guide delay decisions.
The prediction capability enables us to assess whether delaying a function is beneficial, identify the most suitable locality type, and dynamically determine optimal delay thresholds.
As shown in Figure~\ref{fig:design-arch}, \alias integrates \textit{function and node profiling}, \textit{execution time prediction}, and \textit{differentiated delay scheduling}, performed by three components:
1) distributed data agents deployed on each node that continuously collect and push system metrics, warm container status, zygote process information, and package dependency data to the scheduler;
2) an online learning-enabled prediction model that estimates execution times across different nodes to adaptively determine optimal delay thresholds; and
3) a delay scheduler that maintains function characteristics, manages function queuing, and scheduling delay decisions.
The scheduler identifies the most suitable locality strategy for each function and determines whether delaying for locality benefits outweighs costs under current system conditions.

\subsection{Function and Node Profiling}

To enable accurate execution time prediction, \alias implements a comprehensive profiling system that collects both function characteristics and node-level metrics. The goal is to capture all relevant factors that may impact function execution times, including both data locality and infrastructure locality~\cite{mcgrath2017serverless,DBLP:conf/usenix/WangLZRS18,baldini2017serverless}. While node-level metrics are gathered by distributed agents on each worker node and streamed to the scheduler, function-specific metrics are collected by the scheduler itself when functions arrive for scheduling, providing a complete picture for real-time decision making.




For node-level profiling, distributed agents continuously collect metrics that are essential for capturing infrastructure locality potential and its impact on cold start latency~\cite{DBLP:conf/usenix/OakesYZHHAA18,DBLP:journals/corr/abs-1709-10140,DBLP:conf/ic2e/BachiegaSBS18}.
Specifically, we track the number of warm containers available for each function type, which eliminates container initialization overhead when reused~\cite{DBLP:conf/usenix/OakesYZHHAA18,DBLP:conf/usenix/WangLZRS18}. The agents also monitor active zygote processes, which serve as pre-initialized runtime environments that can be quickly forked to create new function instances, substantially reducing startup times. Additionally, we collect package dependency cache status and pre-loaded library information, as these directly affect dependency resolution time, which can dominate execution time for dependency-heavy functions~\cite{DBLP:conf/edge/Elgamal18,DBLP:conf/usenix/ShahradFGCBCLTR20,DBLP:conf/usenix/WangCTWYLDC21}.
Besides container and dependency state metrics, agents also report system resource metrics to provide context on each node's current operational environment~\cite{DBLP:journals/cn/WoodSVY09}.
For example, CPU utilization, load averages, and scheduling latencies help assess potential contention~\cite{DBLP:journals/jnca/ZhangHW16,DBLP:conf/osdi/ShueFS12}. Network metrics, including bandwidth utilization and packet rates, are important for functions that perform significant data transfers or API calls~\cite{DBLP:conf/usenix/WangCTWYLDC21,DBLP:conf/usenix/MahgoubSMKCB21}. Disk I/O metrics help predict performance for storage-intensive operations~\cite{DBLP:conf/usenix/MahgoubSMKCB21,DBLP:conf/sc/ZhaoYLZL21}. These system metrics are collected at regular intervals and streamed to the scheduler.

For function-level characteristics, the scheduler itself collects and maintains relevant information when functions arrive for scheduling. Input data size is recorded as it directly correlates with data transfer times and often with processing requirements~\cite{FunctionBench}. The scheduler maintains a historical record of elapsed times for key execution phases (data transfer, container initialization, and actual function execution) from previous invocations, which serves as valuable training data for our prediction model. For functions that are part of larger workflows, we track dependency information to enable co-location of related functions when beneficial. The scheduler also records required package dependencies for each function, which helps identify nodes with matching pre-loaded dependencies.

All these metrics are continuously updated in the scheduler's state cache, providing a real-time view of both infrastructure conditions and function characteristics.
This comprehensive profiling enables our prediction model to accurately estimate execution times across different nodes, considering both the current system state and function-specific requirements.

\subsection{Function Execution Time Prediction}

Accurate prediction of function execution time is essential for effective delay scheduling.
We model this problem as a regression task, aiming to learn a mapping from the feature space to the function execution time.
In designing the prediction model, we address two key challenges: 1) capturing the diverse factors influencing execution time, such as node status, function characteristics, and locality features; and 2) ensuring the model is both efficient and adaptable to evolving workload patterns and system dynamics.

To this end, we employ online Random Forest Regression (RFR) to build our model~\cite{segal2004machine,DBLP:conf/sc/ZhaoYLZL21}.
RFR provides robust estimates by averaging predictions from multiple decision trees built with bootstrap sampling and random feature selection~\cite{ishwaran2008random}.
Unlike linear regression, which assumes linear relationships between features and execution times, RFR automatically captures complex, non-linear performance dynamics inherent to serverless environments without manual feature engineering.
This involves intricate relationships between system metrics, function characteristics, and performance outcomes.
The choice of RFR over deep learning methods is based on the following considerations: while deep learning may offer higher accuracy, it typically demands more training samples and significantly more computational resources, making it less efficient for real-time environments; RFR is not only simple to implement but also suitable for handling high-dimensional, multi-type continuous variables and is highly robust against overfitting~\cite{smith2013comparison,rodriguez2015machine,gromping2009variable}.
Our experimental results demonstrate that RFR outperforms other representative machine learning algorithms and deep learning models
in terms of prediction accuracy and computational efficiency.

Our prediction model can be formally represented as learning a mapping function $f:\mathcal{X} \rightarrow \mathbb{R}^+$ from the feature space $\mathcal{X}$ to the positive real-valued execution time. Given $n$ observed samples $\{(\mathbf{x}_i, y_i)\}_{i=1}^n$, where $\mathbf{x}_i \in \mathcal{X}$ represents the feature vector and $y_i \in \mathbb{R}^+$ represents the corresponding execution time, our goal is to minimize the prediction error:

\begin{equation}
\min_f \sum_{i=1}^n \mathcal{L}(f(\mathbf{x}_i), y_i) + \lambda \varphi(f)
\end{equation}

\noindent where $\mathcal{L}(\cdot,\cdot)$ is the loss function measuring prediction accuracy, and $\varphi(f)$ is the regularization term with hyperparameter $\lambda$ used to prevent overfitting.
The feature space $\mathcal{X}$ contains a comprehensive feature set that captures system status and function characteristics:

\begin{equation}
\mathbf{x}_i = [\mathbf{s}_i, \mathbf{c}_i, \mathbf{n}_i, \mathbf{d}_i]
\end{equation}

\noindent where $\mathbf{s}_i$, $\mathbf{c}_i$, $\mathbf{n}_i$, and $\mathbf{d}_i$ represent system resource metrics, container status metrics, network metrics, and function characteristic metrics, respectively.
Specifically, $\mathbf{s}_i$ includes metrics such as CPU utilization, memory usage, disk I/O, and system load, which directly affect computing performance and the degree of resource contention.
$\mathbf{c}_i$ captures the status of containers and dependencies, including the number of available warm containers, active zygote processes, and cached dependencies.
These metrics are crucial for assessing infrastructure locality, as they directly impact cold start latency and dependency resolution time.
$\mathbf{n}_i$ records metrics such as network bandwidth utilization, packet rate, and network latency, which are particularly important for data transfer-intensive functions.
$\mathbf{d}_i$ includes function-specific features such as input data size, the number of required dependencies, and historical execution time. These features help the model distinguish between different types of functions and their sensitivity to various types of locality.

Our RFR model constructs an ensemble of decision trees $\{h_k\}_{k=1}^K$, with the final prediction obtained by averaging:

\begin{equation}
f(\mathbf{x}) = \frac{1}{K}\sum_{k=1}^K h_k(\mathbf{x})
\end{equation}

\noindent For each scheduling decision, the model provides an execution time estimate for each candidate node by processing the combined feature vector.
These predictions serve as crucial inputs to the scheduler's delay decision-making process, allowing it to compare expected execution times across different nodes.

To capture the dynamics of the serverless environment, we employ an online learning mechanism to incrementally update the model.
The online RFR model follows a two-phase training approach.
It is initially trained with a basic set of historical execution data (including function/system features and the corresponding function execution times) to establish baseline prediction capabilities.
During system operation, \alias continuously collects new observations to incrementally update the model through online learning~\cite{DBLP:conf/sc/ZhaoYLZL21}.
The update process is automatically triggered when the number of new observations reaches a predefined threshold.
This adaptive approach enables the model to effectively handle the dynamic characteristics of the serverless environment without the overhead of complete retraining.
Together, these techniques, i.e., function and system profiling, ensemble learning, and online model updating, yield accurate execution time predictions, which are used to guide 
\alias's delay scheduling decisions.

\subsection{Differentiated Delay Scheduling}

\begin{algorithm}[t]
\SetAlgoLined
\caption{Differentiated Delay Scheduling}\label{alg:scheduling}
\KwIn{Function $\gamma$, Node set $\mathcal{N}$, SLA threshold $\theta$}
\KwOut{Scheduling decision for function $\gamma$}

\SetKwFunction{FSchedule}{DelayScheduling}
\SetKwFunction{FDelayMonitoring}{DelayMonitoring}

\SetKwProg{Fn}{Function}{:}{}
\Fn{\FSchedule{$\gamma$, $\mathcal{N}$, $\theta$}}{
    $\mathcal{N}_l, \mathcal{N}_f \gets$ ClassifyNodes($\gamma$, $\mathcal{N}$)\;
    
    $\mathcal{T} \gets$ PredictExecutionTime($\gamma$, $\mathcal{N}$)\;

    $\mathcal{T}_l, \mathcal{T}_f \gets \min_{node \in \mathcal{N}_l} \mathcal{T}[node], \min_{node \in \mathcal{N}_f} \mathcal{T}[node]$
    
    \If{$\mathcal{N}_l = \emptyset$ {\rm \textbf{or}} $\mathcal{T}_l \geq \mathcal{T}_f \times \alpha$}{
        $\eta_f \gets \argmin_{node \in \mathcal{N}_f} \mathcal{T}[node]$\;
        Immediately schedule $\gamma$ to $\eta_f$\;
        \Return\;
    }

    
    
    $\eta_l \gets \argmin_{node \in \mathcal{N}_l} \mathcal{T}[node]$\;
    
    \textbf{Start} DelayMonitoring($\gamma$, $\eta_l$, $\mathcal{N}_f$, $\theta$)\;
}

\Fn{\FDelayMonitoring{$\gamma$, $\eta_l$, $\mathcal{N}_f$, $\theta$}}{
    \While{True}{
        \If{$\eta_l$~has~available~resource}{
            Immediately schedule $\gamma$ to $\eta_l$\;
            \Return\;
        }

        $\mathcal{T} \gets$ PredictExecutionTime($\gamma$, $\mathcal{N}_f$)\;
        
        $violation \gets$ True\;
        \ForEach{$node \in \mathcal{N}_f$}{

            $delay \gets \theta \times(1-\beta) - \mathcal{T}[node]$\;
            \If{$delay > 0$}{
                $violation \gets$ False\;
                \textbf{break}\;
            }
        }
        \If{$violation$}{
            $\eta_f \gets \argmin_{node \in \mathcal{N}_f} T[node]$\;
            Immediately schedule $\gamma$ to $\eta_f$\;
            \Return\;
        }
        Sleep($\mathcal{I}$)\;
    }
}
\end{algorithm}

Based on our analysis of locality benefits and SLA constraints, we now present \alias's differentiated delay scheduling (Algorithm~\ref{alg:scheduling}).
The core principle is to evaluate whether delay scheduling is beneficial for each function by considering both locality types and execution time predictions.
Our algorithm combines continuous delay monitoring with event-driven scheduling, which is triggered by function arrivals, resource releases, and SLA violation signals from the delay monitor.
\alias implements these principles through systematic \textit{node classification} and \textit{adaptive delay decision-making} that maximize locality benefits with SLA compliance.


\textbf{Node Classification} serves as a fundamental mechanism in \alias to enable intelligent scheduling decisions by systematically categorizing computational resources according to their current state and locality characteristics.
For each function awaiting scheduling, \alias dynamically classifies available nodes into two distinct categories (line 2), i.e., \textit{local nodes} and \textit{fallback nodes}.
This is done by analyzing both workflow dependencies and infrastructure characteristics.
For data locality, the system examines the function's position within the workflow and identifies nodes that contain intermediate data produced by predecessor functions.
For infrastructure locality, the system queries each node's cached state to determine the availability of warm containers matching the function's runtime requirements and the presence of pre-installed package dependencies.
The fundamental purpose of this step is to distinguish which nodes possess locality characteristics that can enhance function execution performance.
These classifications are inherently function-specific rather than static designations, meaning a node's category may vary considerably across different functions and evolve temporally as underlying node conditions change.
This dynamic classification approach allows the scheduler to make nuanced, context-aware decisions that appropriately balance the potential benefits of locality against the costs of scheduling delays.

\begin{itemize}[noitemsep,leftmargin=5.5mm]
    \item \textit{Local nodes ($\mathcal{N}_l$):} Nodes that currently offer locality benefits for the function through either data locality or infrastructure locality.
    Rather than using discrete binary classifications as simply ``data-local'' or ``infrastructure-local,'' \alias employs a continuous methodology that quantifies the degree of locality advantage each node offers for the current function.
    Specifically, a node delivers data-locality benefits when it contains relevant input data from predecessor functions in the application workflow, particularly for functions with data-intensive workloads.
    Infrastructure-locality benefits manifest when a node hosts warm containers or cached dependencies that can accelerate the startup of heavy-dependency functions.
    A node can simultaneously exhibit both types of locality benefits with varying degrees of effectiveness.
    This continuous approach enables \alias to automatically prioritize nodes based on their locality characteristics.
    In particular, a node's locality status can change dynamically as data are processed or containers expire. Multiple local nodes may exist, and the system only needs to wait for one suitable node to become available.
    

    \item \textit{Fallback nodes ($\mathcal{N}_f$):}
    Nodes with suboptimal locality conditions that provide little performance advantages for function execution.
    They tend to have minimal queueing time (typically less than the \textit{safe delay threshold} to be introduced later), rendering them effectively available for rapid scheduling.
    While they may not offer locality benefits, they serve as reliable fallback options to prevent excessive delays and ensure SLA compliance.
    In line 5-9, if the function lacks local nodes (e.g., if it is newly created), it will be immediately scheduled to the fallback node with the smallest predicted execution time.
    A node may transition between local and fallback status as its resource availability and locality characteristics change over time.

\end{itemize}

\textbf{Adaptive Delay Decision-making} comprises the following two key steps to schedule a function:

\begin{enumerate}[noitemsep,leftmargin=5.5mm]
    \item \textit{Locality target selection:} \alias determines the optimal locality target by comparing predicted execution times across all available local nodes for the function, selecting the one that yields the lowest predicted execution time.

    \item \textit{Delay benefit validation:} The system validates whether delaying is worthwhile by comparing the predicted execution time on the target local node ($\mathcal{T}_l$) with the minimum predicted time on fallback nodes ($\mathcal{T}_f$). Delay scheduling proceeds (line 10-11) only if $\mathcal{T}_l < \mathcal{T}_f \times \alpha$, where $\alpha~(\alpha<1)$ is a configurable threshold balancing locality benefits against delays (line 5).
    This condition ensures that delay scheduling is only applied when the execution time on a local node is significantly better than on fallback nodes, providing a quantifiable performance advantage that justifies the additional waiting time. 
    Without this validation, the system might delay execution for minimal performance gains, potentially wasting time that could be better spent executing the function immediately on a fallback node.
\end{enumerate}

When a function enters delay state, a dedicated thread monitors the scheduling decision every $\mathcal{I}$ ms (default \SI{100}{ms}, line 32) until either a local node becomes available (line 14-17), or the safe delay threshold approaches zero (line 20-26).
This safe delay threshold represents the maximum time a function can safely wait for a local node without risking SLA violation.
It can be determined by subtracting the predicted execution time from the product of the SLA and a safety margin factor $(1-\beta)$ (line 18). 
Both the monitoring interval $\mathcal{I}$ and safety margin $\beta$ are configurable parameters: shorter intervals provide more responsive scheduling at increased system overhead, while larger safety margins better protect against SLA violations at the cost of reduced locality opportunities.
This approach ensures dynamic delay decisions based on both function characteristics and current system conditions while maintaining SLA compliance through continuous monitoring and adaptive thresholds.

\section{Implementation}

\alias is implemented as a distributed system on OpenLambda~\cite{DBLP:conf/usenix/OakesYZHHAA18,hendrickson2016serverless}.
We leverage OpenLambda's inherent capability to distribute serverless functions across a cluster of worker nodes while extending it with locality-aware and delay scheduling capabilities.
The system architecture consists of four main components, i.e., a \textit{gateway}, a \textit{scheduler}, a \textit{prediction model}, and \textit{worker agents}.
\alias uses different programming languages for its components to optimize their respective performance.
The gateway handles function invocation requests and load balancing, serving as the primary entry point for function invocations.
The scheduler operates centrally and makes global scheduling decisions based on cluster-wide information.
The prediction model runs as a separate service to provide execution time forecasts.
Worker agents on each worker node execute functions and continuously stream local metrics (CPU, memory, network I/O, container status) to the scheduler at configurable intervals (default \SI{50}{ms}), maintaining an up-to-date view of cluster-wide system state.



Our architecture emphasizes efficient data collection and processing through several key mechanisms that improve distributed system scalability.
Worker agents extend OpenLambda with lightweight monitoring capabilities for serverless-specific metrics like warm container counts and package information, which are cached in memory for rapid access.
The system uses gRPC with server-side streaming, enabling worker agents to push state updates to the scheduler without polling overhead, which ensures efficient cluster-wide coordination.
The scheduler maintains the system states in its local cache and performs its scheduling algorithms exclusively on this local cache, eliminating network round-trips during critical decision-making processes.
Moreover, function execution predictions are requested from the model service via gRPC only when necessary.
To enhance responsiveness, time-consuming operations are moved off the critical path of function execution and scheduling decisions.
Specifically, delay monitoring runs in separate goroutines to prevent blocking the main scheduling loop, while model inference and metric collection operate asynchronously.
These architectural decisions enable \alias to maintain low-latency scheduling decisions even at high concurrency levels.


\section{Experimental Evaluation}
In this section, we conduct extensive experiments to evaluate the performance of \alias.
Our evaluation aims to answer the following research questions:

\begin{itemize}[noitemsep,leftmargin=5.5mm]
    \item \textbf{RQ1}: How effective is \alias's differential delay scheduling in improving function execution performance compared to baseline methods?
    \item \textbf{RQ2}: How do the performance advantages of \alias change as the concurrency level increases (which lead to heavy node loads and resource contention)?
    \item \textbf{RQ3}: How does \alias's prediction model perform in terms of accuracy and adaptability. 
    Additionally, how does the scheduler itself perform regarding overhead and scalability?
\end{itemize}

\textbf{Experimental Setup:} We implement and deploy \alias on a cluster of ten nodes, each equipped with an Intel(R) Xeon(R) Gold 5318Y CPU @ 2.10GHz (16 cores), 8GB RAM, 500GB storage, and connected via 1Gbps networking. The experiments are conducted using OpenLambda~\cite{DBLP:conf/usenix/OakesYZHHAA18,hendrickson2016serverless} as the underlying serverless platform.
We compare the following scheduling strategies:

\begin{itemize}[noitemsep,leftmargin=5.5mm]
    \item \textit{Basic Scheduling (BS)}: A conventional round-robin scheduling approach that distributes function invocations across available nodes without considering locality characteristics. This strategy represents typical serverless platforms that prioritize load balancing and immediate execution over locality benefits.
    
    \item \textit{Naive Locality Scheduling (NLS)}: A locality-aware scheduling approach that considers both data and infrastructure locality when making placement decisions.
    It executes functions immediately on local nodes without delay, but falls back to nodes without locality when local resources are unavailable.
    This strategy attempts to leverage locality benefits when possible but does not wait for optimal nodes to become available.

    \item \textit{Rule-based Delay Scheduling (RDS)}: The same strategy employed in our empirical study (Section~\ref{sec:empirical_study}), which follows the rules in~\cite{DBLP:conf/eurosys/ZahariaBSESS10} to determine the delay thresholds.
    
    \item \textit{XFaaS}: The scheduling approach of XFaaS~\cite{DBLP:conf/osdi/SchusterDVDHGKMAA21} focuses on infrastructure locality optimization.
    It partitions functions and workers into locality groups to maximize warm container reuse and cached dependency benefits.
    XFaaS employs a heuristic delay scheduling mechanism specifically for functions marked as non-critical (requiring manual annotations), allowing these functions to be delayed during high-load periods to preserve resources for critical functions.
    For each application, we annotate user-facing and latency-sensitive workflow stages as critical and tune the remaining non-critical annotations to minimize mean end-to-end latency without violating the SLA.
    
    \item \textit{AQUATOPE}: A QoS-and-uncertainty-aware resource management system~\cite{DBLP:conf/asplos/ZhouZD23} that improves serverless performance through dynamic pre-warmed container pool management. AQUATOPE employs a hybrid Bayesian neural network to predict future function invocation rates and proactively adjusts the number of warm containers to eliminate cold starts.
    For each application, we conduct a training phase, allowing its Bayesian model and Bayesian optimization engine to process sufficient workflow executions until the recommended resource configurations and end-to-end latency stabilize.
    Icebreaker~\cite{DBLP:conf/asplos/RoyPT22} follows a similar principle to reduce cold start latency by predicting function invocation and proactively warming containers.

\end{itemize}

Beyond the selected baselines, there are many other approaches proposed for serverless performance optimization.
However, these works focus on different layers of the system stack that are orthogonal to \alias.
For instance, approaches targeting cold start reduction through custom runtimes~\cite{DBLP:conf/usenix/OakesYZHHAA18,DBLP:conf/asplos/YuRFTLZWP24} or kernel-level modifications~\cite{shillaker2020faasm,Catalyzer}, efficient data transfer mechanisms~\cite{DBLP:conf/usenix/MahgoubSMKCB21,Akkus_Chen_Rimac_Stein_Satzke_Beck_Aditya_Hilt_2018}, and workflow orchestration systems~\cite{DBLP:conf/eurosys/AbdiGLFCGBBF23} represent infrastructure-level optimizations that could potentially be combined with \alias's scheduling algorithms.
Moreover, our baselines capture the fundamental scheduling principles employed by many existing approaches in both serverless computing and distributed systems, e.g., Palette~\cite{DBLP:conf/eurosys/AbdiGLFCGBBF23}, Netherite~\cite{DBLP:journals/pvldb/BurckhardtCGJKM22}.
Therefore, our baseline selection follows established practices in serverless scheduling evaluation and enables focused assessment of our algorithmic contributions.

We evaluate these strategies using the same four representative serverless applications as in the pre-experiments (Section~\ref{sec:empirical_study}).
Each experiment is repeated multiple times to ensure statistical significance.
We measure execution times at various percentiles to understand both average-case and tail-latency performance.

\subsection{RQ1: The Effectiveness of \alias}

\begin{figure}[t]
    \centering
    \includegraphics[width=0.92\columnwidth]{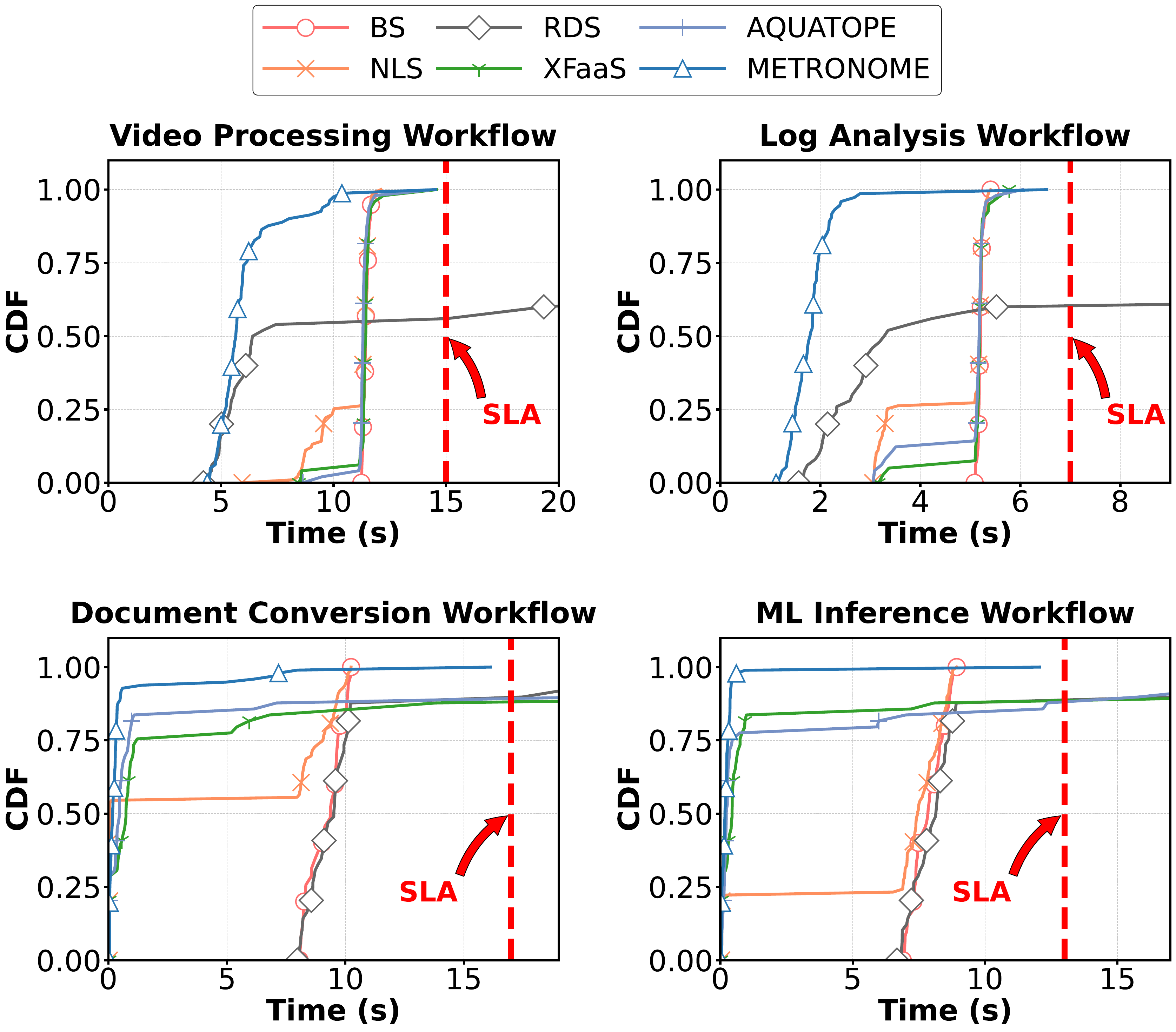}
    \vspace{-8pt}
    \caption{CDF of function execution time}
    \Description{Line chart showing cumulative distribution functions of function execution time under different scheduling strategies, where curves closer to the upper-left corner represent lower latencies and better overall performance.}
    \label{fig:perf_comparison}
    \vspace{-8pt}
\end{figure}

Figure~\ref{fig:perf_comparison} presents the CDFs of application execution times under different approaches. The results show that \alias achieves substantial performance improvements across various function types, with patterns varying according to function characteristics.

For applications involving substantial data transfer, \alias yields considerable improvements in both average and median execution times.
Specifically, in video processing, it reduces mean execution time by 47.1\% and 90th percentile latency by 31.2\% compared to BS,
with similar gains over NLS, XFaaS and AQUATOPE.
When compared to RDS, the mean and 90th percentile reductions are 74.8\% and 87.2\%, respectively.
In log analysis, \alias demonstrates even more significant gains compared to all baselines but RDS, i.e., with approximately 64\% reduction in mean and 58\% improvement in 90th percentile.
When compared to RDS, both metrics realize over 90\% enhancement.
These advances can be attributed to \alias's prioritization of data locality, reducing data transfer overhead between function stages.
For dependency-heavy applications, \alias's infrastructure locality optimization is highly effective.
Document conversion achieves around 93\% and 95\% reductions in mean execution time and 90th percentile latency, respectively, across most baselines. In particular, \alias demonstrates an 83.9\% reduction in mean execution time compared to AQUATOPE.
ML inference demonstrates similar benefits.

Overall, the enhancements over RDS are more substantial, which highlights the limitations of traditional rule-based delay scheduling approaches in serverless environments.
For document conversion and ML inference, RDS performs even worse than NLS as the delay overhead outweighs any potential data locality benefits.
This is because NLS considers infrastructure locality while RDS does not, which confirms the benefits of this new dimension of locality as revealed in our study.
The improvements over XFaaS demonstrate the value of an automated and intelligent delay scheduler.
XFaaS's delay scheduling requires manual annotation of functions as 
``non-critical'' and heuristic, time-shifting rules to defer these 
functions to off-peak hours (e.g., overnight).
The lack of data locality consideration also limits its effectiveness for data-intensive workflows.
The gains over AQUATOPE can be attributed to its aggressive warm container provisioning strategy. While reducing cold starts, it can lead to resource contention and interference between co-located functions, ultimately degrading end-to-end performance.
Additionally, AQUATOPE's complex Bayesian optimization process introduces significant computational overhead in the decision-making process. Data locality is also not considered in AQUATOPE.

Despite these improvements, we observe some increases in tail latencies compared to BS and NLS, which, however, are minimal and affect only a small percentage of executions.
Nevertheless, \alias still delivers positive improvements in 99th percentile latency over BS and NLS, i.e., roughly 6\% for video processing, 30\% for log analysis, 19\% for document conversion, and 79\% for ML inference.
Gains are even more pronounced relative to RDS, XFaaS, and AQUATOPE.
This demonstrates that even in extreme cases, \alias maintains performance advantages.
Furthermore, our adaptive delay mechanism successfully prevents any SLA violations across all applications, demonstrating an effective balance between waiting for optimal nodes and meeting latency requirements.


\subsection{RQ2: Scalability Evaluation}

\begin{figure}[t]
    \centering
    \includegraphics[width=0.93\columnwidth]{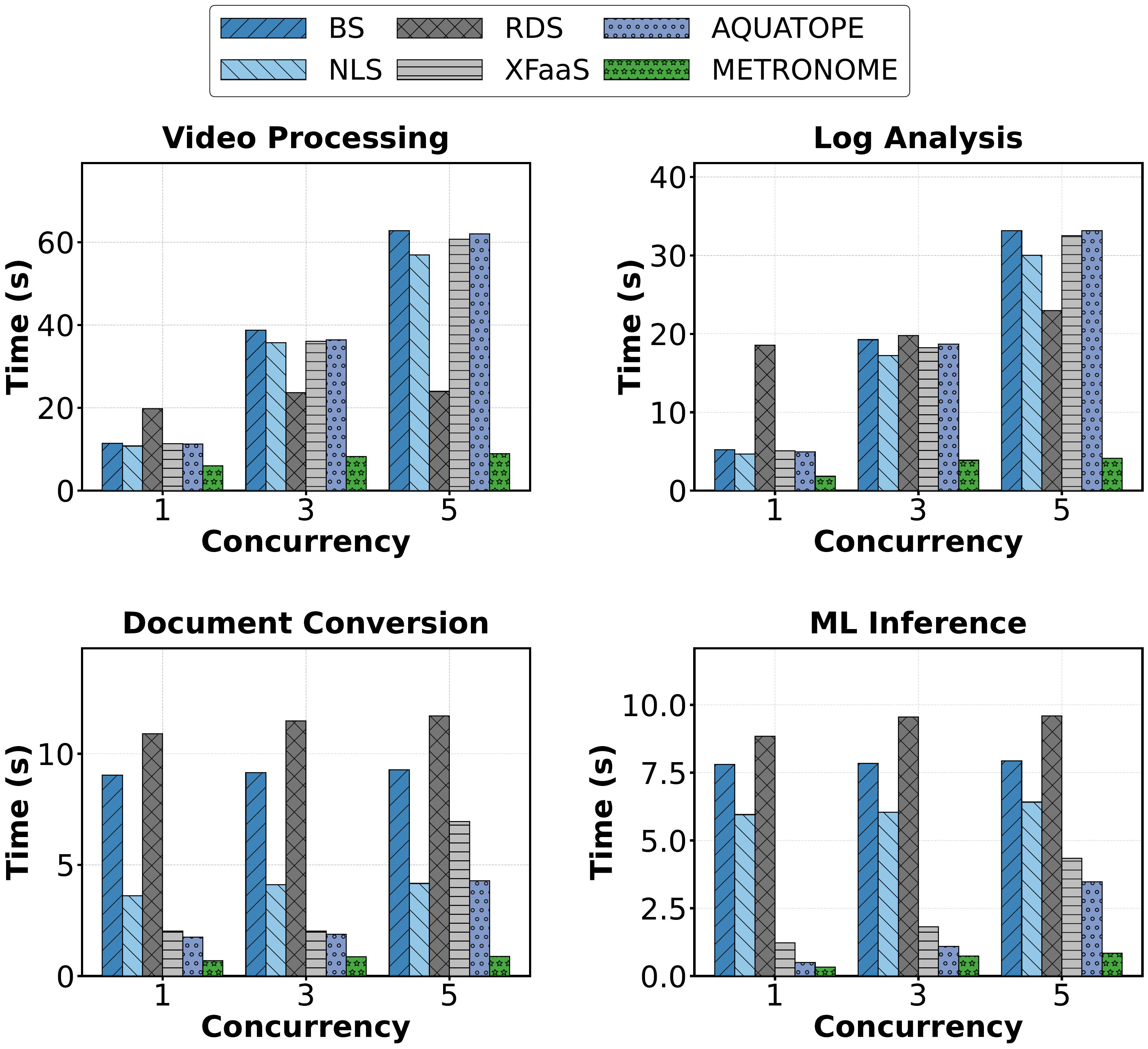}
    \caption{Mean execution times under diff. concurrency levels}
    \Description{Grouped bar chart summarizing mean workflow execution time at different concurrency levels for each application and scheduling strategy, where taller bars indicate slower execution and poorer scalability.}
    \label{fig:scalability_comparison}
    \vspace{-3pt}
\end{figure}

We evaluate \alias's performance under varying concurrency levels (1, 3, and 5 concurrent executions) to assess its scalability. This evaluation is particularly important because in production, serverless functions are typically invoked concurrently by multiple users, creating resource contention and potentially affecting locality benefits. Figure~\ref{fig:scalability_comparison} presents the mean execution times across these concurrency levels for all applications, providing insights into how \alias performs under realistic multi-user scenarios.

The results reveal distinct scaling patterns across application types. For data-intensive applications (video processing and log analysis), both BS and NLS exhibit substantial performance degradation as concurrency increases, likely due to increased network contention when transferring intermediate data between workflow stages.
RDS shows some improvements at higher concurrency levels.
However, its rule-based delay threshold mechanism leads to significant performance degradation in certain scenarios, even worse than both BS and NLS due to excessive tail latencies.
XFaaS and AQUATOPE exhibit mixed scaling behavior.
They achieve moderate improvements for dependency-heavy applications but suffer noticeable performance degradation in data-intensive functions.
This is due to increased resource contention when multiple instances compete for limited bandwidth.
In contrast, \alias demonstrates significantly better scalability with only modest execution time increases at higher concurrency levels. This improved scalability can be attributed to \alias's ability to maintain data locality through its differentiated delay scheduling mechanism, effectively reducing network contention under higher load.

For dependency-heavy applications like document conversion and ML inference, the impact of increased concurrency is less pronounced across all strategies. Their primary overhead stems from package initialization and dependency loading rather than data transfer. Once dependencies are cached, they can be reused across multiple function instances with minimal additional overhead. 
RDS performs significantly worse in this case because it only focuses on data locality.
XFaaS exhibits scalability limitations under high concurrency due to its static locality grouping and heuristic scheduling strategy, which can lead to load imbalances as workload patterns evolve.
AQUATOPE's performance degrades under increased concurrency as its resource allocation operates at coarse temporal granularity, while its container-centric optimization approach may not address all locality requirements for diverse workload types.
Additionally, AQUATOPE exclusively focuses on infrastructure locality and neglects data locality consideration, which becomes particularly critical for data-intensive workflows under high load.


\subsection{RQ3: Model and Scheduler Performance}

\begin{figure}
    \centering
    \includegraphics[width=0.8\columnwidth]{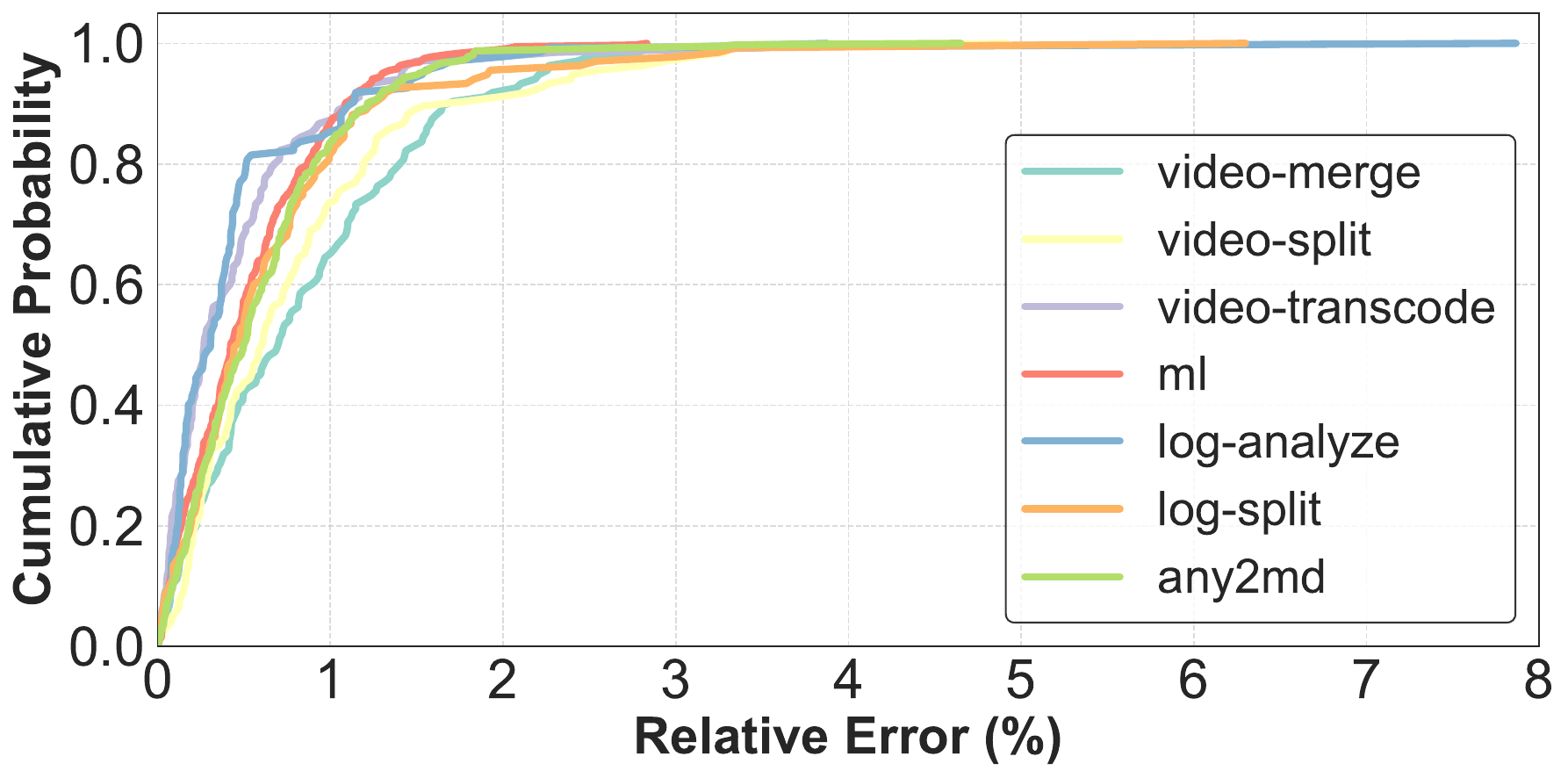}
    \vspace{-6pt}
    \caption{CDF of prediction errors across different functions}
    \Description{Line chart plotting cumulative distribution functions of relative prediction error for multiple functions, with the x-axis showing error magnitude and the y-axis the fraction of samples whose error is below that value.}
    \vspace{-6pt}
    \label{fig:prediction_error}
\end{figure}

We evaluate our prediction model's effectiveness across various function types. Figure~\ref{fig:prediction_error} illustrates the CDF of relative prediction errors, demonstrating high accuracy with a median error of 0.44\% and a 90th percentile error of 1.28\%. The model maintains reasonable accuracy even at the tail, with 95th and 99th percentile errors of 1.63\% and 2.98\%, respectively.


To assess practical feasibility, we analyze the model's computational overhead by calculating the prediction latency distribution across 550 samples.
Our model demonstrates efficient inference with mean latencies of \SI{37.3}{ms}, respectively, while maintaining reasonable tail performance with 95th and 99th percentile latencies of \SI{45}{ms} and \SI{54}{ms}.
These values suggest negligible prediction overhead compared to typical function execution times, making real-time integration feasible.

\begin{figure}
    \centering
    \includegraphics[width=0.83\columnwidth]{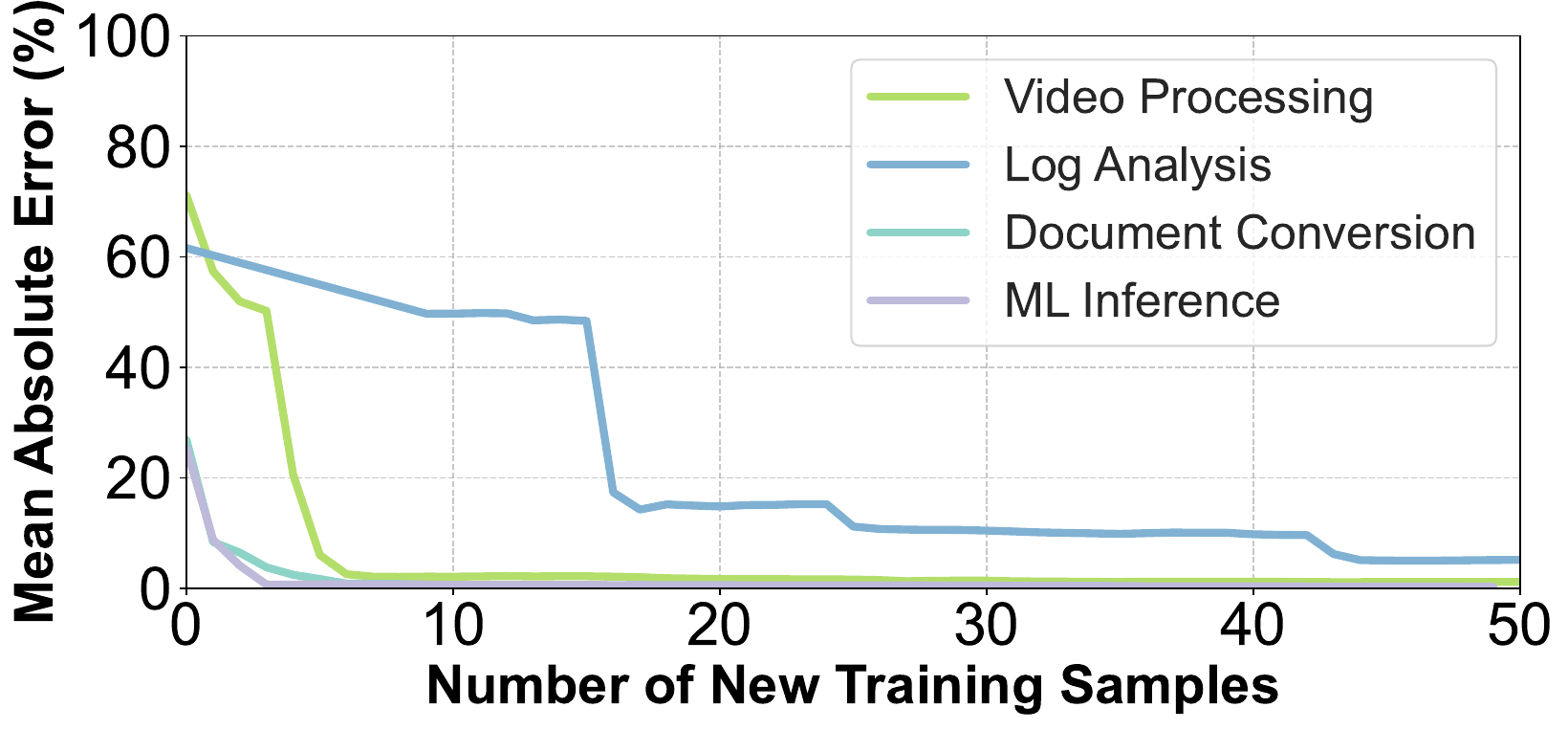}
    \vspace{-6pt}
    \caption{Model prediction error recovery with new samples}
    \Description{Multiple recovery curves showing prediction error decreasing as more samples are observed for each application, with lines dropping rapidly at first and then flattening as the model adapts to the workload.}
    \vspace{-6pt}
    \label{fig:model_recovery}
\end{figure}


We further examine the model's adaptability to emerging workload patterns.
Figure~\ref{fig:model_recovery} depicts error recovery curves as the model encounters previously unseen function characteristics.
The model demonstrates rapid adaptation capabilities. 
Specifically, in data-intensive applications, video processing errors drop below 10\% after just 5 samples, while log analysis requires approximately 30 samples to achieve similar accuracy.
Dependency-heavy applications converge even faster, with document conversion and ML Inference stabilizing below 5\% error after only 2-3 samples.
All application types eventually stabilize at 1-5\% error rates after sufficient training samples.
This differentiated adaptation pattern aligns with our earlier findings on application characteristics and demonstrates that our online learning mechanism effectively adapts to diverse workload patterns without extensive retraining.

\begin{figure}
\centering
\begin{subfigure}[b]{0.48\columnwidth}
    \includegraphics[width=\textwidth]{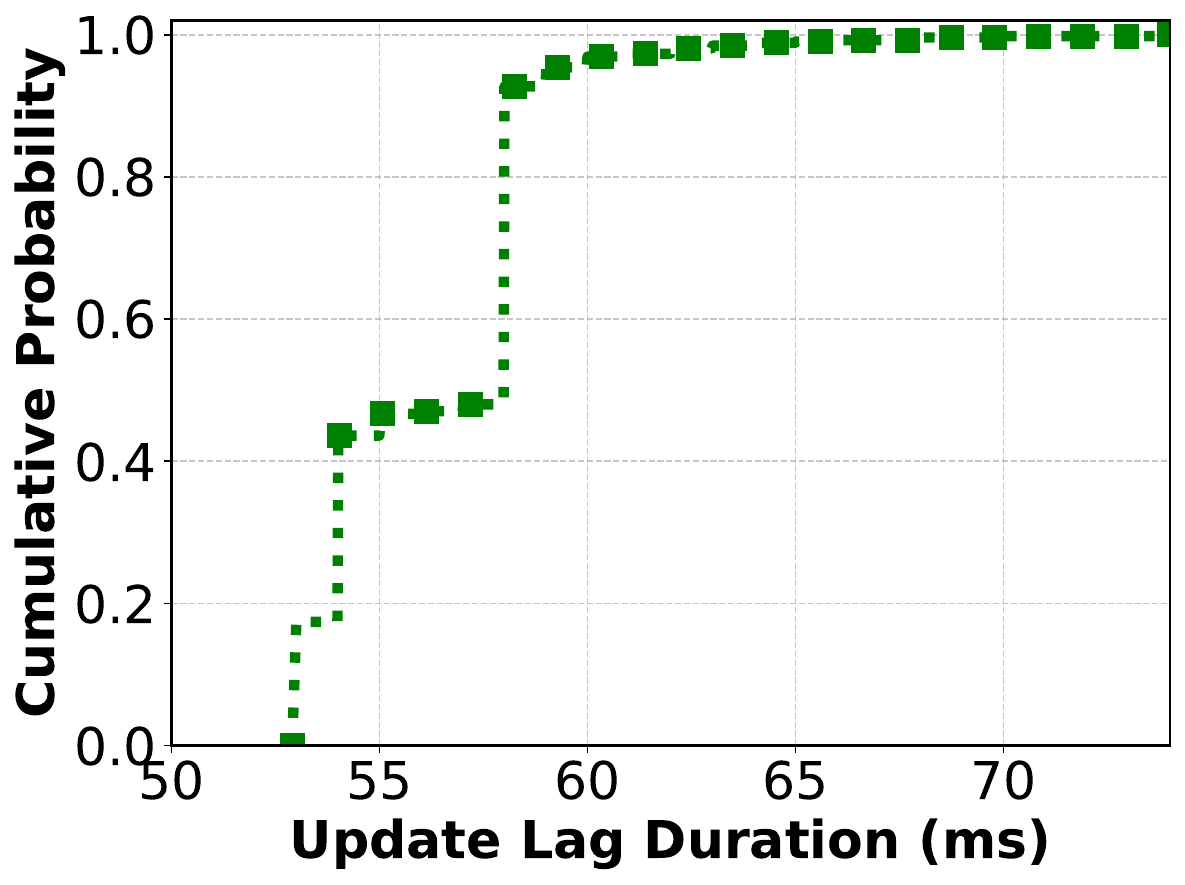}
    \caption{CDF of system metric update lag duration}
    \label{fig:update_lag}
\end{subfigure}
\hfill
\begin{subfigure}[b]{0.48\columnwidth}
    \includegraphics[width=\textwidth]{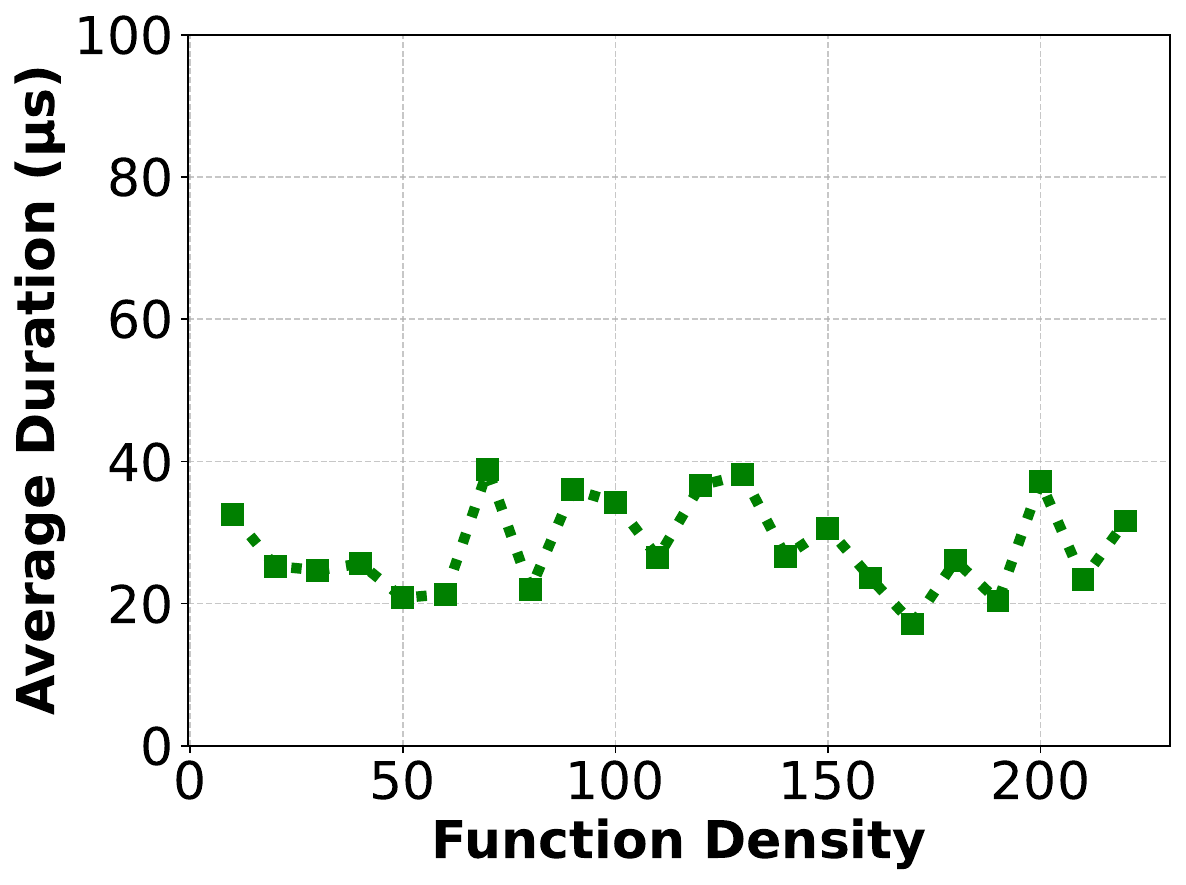}
    \caption{\alias's scheduling decision latency}
    \label{fig:scheduling_lag}
\end{subfigure}
\vspace{-4pt}
\caption{Performance of the delay scheduler}
\Description{Composite figure with two side-by-side subplots: the left subplot shows the cumulative distribution of system metric update lag, and the right subplot shows scheduling decision latency, together illustrating that both metrics remain low even under load.}
\vspace{-6pt}
\label{fig:scheduler_perf}
\end{figure}

For scheduler performance, we evaluate its overhead by analyzing system metrics collection update lag and scheduling decision latency, both critical for system's real-time capabilities.
Figure~\ref{fig:update_lag} presents the CDF of update lag durations for system metrics collection. Our implementation uses gRPC server-side streaming to push metrics updates every \SI{50}{ms}. Results show consistent metrics updates with approximately 45\% completing within \SI{55}{ms} and another 45\% between 55-\SI{60}{ms}. The 95th percentile lag remains at \SI{60}{ms}, ensuring the scheduler maintains an up-to-date system state view without polling.
Similarly, scheduling decision latency (Figure~\ref{fig:scheduling_lag}) is remarkably low.
By moving network operations off the critical path, our scheduler achieves microsecond-level efficiency with 45\% of decisions made within 55$\mu$s and most others between 55-60$\mu$s. The 95th percentile latency of 60$\mu$s indicates consistently fast and predictable core scheduling logic even in worst-case scenarios.
These metrics demonstrate that our implementation achieves minimal-overhead scheduling while maintaining sophisticated locality-aware placement capabilities. The sub-60$\mu$s latency represents less than 0.1\% of typical function execution times (hundreds of milliseconds to seconds), ensuring our locality-aware scheduling operates efficiently without becoming a system bottleneck.

\section{Related Work}

\textbf{Scheduling for Serverless Computing:}
Scheduling optimization in serverless computing has attracted significant attention.
For example, Hermod~\cite{DBLP:conf/cloud/KaffesYK22} and Palette~\cite{DBLP:conf/eurosys/AbdiGLFCGBBF23} investigate key scheduling decisions around timing, node selection, and intra-node task scheduling.
For QoS-aware resource management, FaaSConf~\cite{DBLP:conf/kbse/Wang0D0YHH24} and scalable serverless architectures~\cite{Ji2025Designing} focus on resource configuration optimization.
Workload characterization studies~\cite{DBLP:conf/usenix/ShahradFGCBCLTR20,HuaweiWorkload} have revealed significant heterogeneity in serverless function characteristics.
General datacenter scheduling approaches from systems like Hawk~\cite{Delgado_Dinu_Kermarrec_Zwaenepoel_2015}, Eagle~\cite{Delgado_Didona_Dinu_Zwaenepoel_2016},
Sparrow~\cite{Ousterhout_Wendell_Zaharia_Stoica_2013}, and Pigeon~\cite{wang2019pigeon}
have proposed various techniques to balance efficiency and scalability in distributed computing environments.
Building upon them, \alias focuses on adaptive scheduling decisions that consider function execution characteristics and system state, enabling more efficient resource utilization while maintaining SLA compliance.

Another closely related domain is microservice scheduling~\cite{DBLP:conf/kbse/MengSTPY23}.
However, microservices typically operate as long-running services with persistent state and continuous execution, whereas serverless functions are ephemeral, stateless, and event-driven.
These fundamental differences highlight the need for specialized scheduling mechanisms that account for the unique characteristics of serverless computing, such as cold start latency, container warm-up overhead, and short function lifetime.


\textbf{Locality Optimization:} Locality optimization in serverless computing spans two dimensions: data and infrastructure. 
For data locality, Palette~\cite{DBLP:conf/eurosys/AbdiGLFCGBBF23} improves the locality between functions in workflows through scheduling optimization. SONIC~\cite{DBLP:conf/usenix/MahgoubSMKCB21} explores efficient data passing mechanisms for chained serverless applications. Pocket~\cite{Klimovic_Wang_Stuedi_Trivedi_Pfefferle_Kozyrakis_2018}, SAND~\cite{Akkus_Chen_Rimac_Stein_Satzke_Beck_Aditya_Hilt_2018}, and the work by Pu et al.~\cite{Pu_Venkataraman_Stoica_2019} address efficient data handling.
Netherite~\cite{DBLP:journals/pvldb/BurckhardtCGJKM22} addresses data locality by co-locating functions and their data within the same partitions.
For infrastructure locality, SOCK~\cite{DBLP:conf/usenix/OakesYZHHAA18} introduces serverless-optimized containers, RainbowCake~\cite{DBLP:conf/asplos/YuRFTLZWP24} proposes hierarchical container caching, Pagurus~\cite{DBLP:conf/usenix/LiG0CXZSMY0G22} explores container-sharing approaches, and FaaSLight~\cite{DBLP:journals/tosem/LiuWCLCLWJ23} focuses on cold-start latency optimization.
XFaaS~\cite{DBLP:conf/osdi/SchusterDVDHGKMAA21} groups functions and workers to maximize warm container reuse benefits while employing delay scheduling for non-critical functions.
AQUATOPE~\cite{DBLP:conf/asplos/ZhouZD23} addresses infrastructure locality through dynamic pre-warmed container pool management, using predictive models to maintain optimal container pool sizes.
Icebreaker~\cite{DBLP:conf/asplos/RoyPT22} is similar to AQUATOPE, which also considers server heterogeneity to balance performance and cost.
\alias uniquely combines both data and infrastructure locality considerations into a unified scheduling framework that automatically determines the most beneficial locality type for each function.

\textbf{Performance Prediction:} Accurate performance prediction is essential for informed scheduling.
Zhao et al.~\cite{DBLP:conf/sc/ZhaoYLZL21} studied inter-function interference in serverless environments and proposed a machine learning approach to predict interference levels. Jiagu~\cite{DBLP:conf/usenix/LiuY0XZFL024} decouples prediction and decision making to
achieve efficient and fast scheduling.
Recent work has also focused on performance variance analysis~\cite{DBLP:journals/ese/WenCSW25} and performance testing methodologies~\cite{DBLP:journals/tosem/WenCZSPZWL25} for serverless computing.
For the general cloud computing field, Quasar~\cite{delimitrou2014quasar}, Paragon~\cite{delimitrou2013qos}, Cuanta~\cite{govindan2011cuanta}, ASM~\cite{subramanian2015application}, Ernest~\cite{venkataraman2016ernest}, and other approaches~\cite{zhao2015predicting,mishra2017esp} employ various techniques to predict performance under different resource allocations. 
\alias extends this line of research by introducing a serverless-specific prediction model that incorporates function features and system metrics. Our approach leverages random forest regression for its computational efficiency and interpretability, while employing online learning to adapt to evolving workload patterns. The prediction results are directly integrated into our delay scheduling framework, enabling effective balancing of locality benefits with SLA compliance.

\section{Conclusion}

This paper presents \alias, a differentiated delay scheduling framework for serverless computing that addresses the varying locality requirements across different functions. 
Our key contributions include: 1) a differentiated scheduling approach that automatically identifies and prioritizes the most appropriate locality type for each function; 2) an adaptive delay threshold mechanism with online learning capabilities that balances locality benefits with SLA compliance; and 3) an efficient implementation architecture that maintains low-latency scheduling decisions while collecting comprehensive system metrics.
Our experimental evaluation demonstrates that \alias achieves significant performance improvements, i.e., up to 64.88\% reduction in mean execution time for data-intensive workflows and up to 95.83\% for dependency-heavy functions, while successfully preventing SLA violations.
The system maintains these performance advantages under increased concurrency levels where traditional approaches experience substantial degradation.
The prediction model demonstrates high accuracy and rapid adaptation to new workload patterns, while the scheduling mechanism operates with negligible overhead.

\begin{acks}
This work was supported in part by the National Key Research and Development Program of China (No. 2023YFB2703600) and the National Natural Science Foundation of China (No. 62402536).
\end{acks}

\bibliographystyle{ACM-Reference-Format}
\balance
\bibliography{citations}


\end{document}